\documentclass{article}
\title{Deep Learning and Model Independence\footnote{This work was funded by the Deutsche Forschungsgemeinschaft (DFG).}}
\author{Martin King\footnote{m.king@lmu.de. Munich Center for Mathematical Philosophy, Ludwigs-Maximilian University Munich}
}
\date{Novemeber 1, 2024}
\usepackage{graphicx}
\usepackage{pgfplots}
\usepackage{enumerate}
\graphicspath{ {../../images/} }
\usepackage{xcolor}
\usepackage{amsmath,amssymb}
\pgfplotsset{compat=1.16}
\usepackage{natbib}

\begin{document}

\maketitle

\begin{abstract}
Despite probing physics at unprecedented energies at the Large Hadron Collider, the Standard Model remains empirically adequate, though incomplete. The lack of evidence in favor of any new physics models means that the search for new physics beyond the Standard Model (BSM) is wide open, with no direction clearly more promising than any other. This marks a turn towards what can be called `model-independent' methods---strategies that reduce the influence of modelling assumptions by performing minimally-biased precision measurements, using effective field theories, or using Deep Learning methods (DL). In this paper, I present the novel and promising uses of DL as a primary tool in high energy physics research, highlighting the use of autoencoder networks and unsupervised learning methods. I advocate for the importance and usefulness of the concept of model independence and propose a definition that recognizes that independence of models is not absolute, but comes in degrees. 
\end{abstract}

\section{Introduction}

Since the 2012 discovery of the Higgs boson at the Large Hadron Collider (LHC), a shift has been underway in the methodology of high energy physics (HEP).\footnote{I will use the term `high energy physics' to stress that this is not so much about the theory of particle physics, but about the study of high energy particle interactions, often conducted at accelerator facilities.} 
The successful Standard Model (SM) research programme dominated the theoretical and experimental sides of HEP research over the past 50 years.
But since the discovery of the final elusive particle, the research programme is coming to an end and no new SM particles remain to be discovered. 
There have been no hints of any additional new physics in the data and the SM remains valid up to the precision and energies to which it has been tested. 
Having no statistically significant anomalies surely speaks in favor of the merits of the SM as a theory, but physicists will not rest contented, as they are convinced (for good reasons) that there is more physics to be discovered.
But the success of the SM and the lack of evidence in favor of new physics models means that the search for new physics is wide open, with no direction clearly more promising than any other. 
Physicists are thus left questioning previously unchallenged assumptions, searching for radically new ideas for guidance, and performing analyses reaching outside the traditional toolkit.
This marks a turn towards what are called `model-independent' methods---strategies that reduce the influence of modelling assumptions by performing minimally-biased precision measurements, using effective field theories (EFTs) to parametrize deviations, or using machine learning (ML) and deep learning methods (DL) to process and analyze the data in new ways.

These changes are of significant philosophical interest for a number of reasons. 
First, they have direct bearing on methodological issues in the philosophy of science: the shift away from model-based searches and towards model-independent approaches suggests a shift from hypothesis testing towards more inductivist methods.
Second, the re-evaluation of long-standing guiding principles, such as naturalness, which was critical in the popularity of supersymmetry, demands a fresh look at the role of scientific values in models of new physics.
Lastly, DL is rife with concerns about the reliability of knowledge production. 
This is highlighted in much of the work being done on explainable artificial intelligence (AI) \citep[such as e.g.][]{buckner2019,hancoxli2020,mittelstadt2019,Sullivan2019} and case studies that have revealed the complexity involved with using DL as a primary tool in physics research \citep{karagiorgi2021machine,Plehn:2022ftl}.
The shift towards model independence is a change worth treating in detail as we may be entering a new era of particle physics. 
Due to the fact that new accelerators would be required to create new heavy fundamental particles, it could be decades or more before we make progress in confirming new theories of physics, until which time, if it comes, we will likely witness the dominance of model-independent methods. 
Of course, model testing and pure theoretical research continues and one never knows when or where a breakthrough may be made.

The term `model independence' is invoked to describe a wide variety of methods and approaches. 
Some prefer to use the term `model agnostic', citing the reason that nothing is completely free of models, e.g. \citep{Dillon:2023zac}.  
This paper aims to characterize the concept of model independence in a way that is faithful to the use in HEP, but is also fruitful in making sense of the different views by allowing that the concept comes in degrees. 
Model independence is a spectrum and a strategy, analysis, or method can be independent from models in different ways and to different degrees. 
The goal of the paper is to describe this new model-independent methodology for scientific discovery, its aims, prospects, drawbacks, and open problems by looking at how it can be achieved through DL.

DL in HEP makes a prime case study for this concept since there is an aim to let algorithms search through the data for indications of new physics without specifying what this new physics is or how the network should find it.
The case study focuses on anomaly detection methods that aim to discover deviations from a well-defined set of background data, in this case the SM.  
Leaving searches open in this way greatly increases the difficulty of the search and until the advent of modern ML methods (which is to say, DL) in the last decade, it was essentially intractable. 
Many difficulties remain and the prospects for the discovery of new physics may still be slim, but DL in HEP is still in its infancy and already represents a powerful new tool for data analysis.

The following section (Section~\ref{sec:rmi}) will describe the current state of affairs in HEP, in particular the shift to model independence in searches for new physics. 
Here, I will also argue that model independence should be seen as a spectrum; methods are independent from models in different ways and to various degrees, but that the term nonetheless represents an important and meaningful concept. 
I will then introduce DL and describe some of the ways in which it is being employed in HEP in Section~\ref{sec:dl}, focusing on the use of autoencoder networks (AENs) and unsupervised learning methods\footnote{I use this term for methods employed on unlabeled data generally (see also Sec.~\ref{sub:dl_basic}).}.
Finally, I will present some concerns about the model independence of these methods in Section~\ref{sec:lessons}.
Ultimately, I will argue that the compromises involved are not fatal to the aim of model independence and the benefits of the DL programme in HEP are well worth exploiting for the model independence they bring into a broad search strategy.

\section{The Rise of Model Independence}	\label{sec:rmi}

\subsection{Finding New Physics}		\label{sub:newphys}

There are two typical kinds of searches for new physics: those looking for deviations from SM predictions using Monte Carlo sampling, and those based on searches for signal events from the predictions of particular BSM models.
To date, there have been no statistically significant deviations from the SM's predictions and the predictions of BSM models are not being confirmed.
This speaks very highly to the success of the SM, but it is frustrating to physicists who are looking for new physics. 
With no new higher energy accelerator forthcoming and as no new physics has been revealed in the LHC data, hopes for the discovery of a more fundamental and more encompassing theory are becoming slim. 
As a result, physicists have had to get very clever with what is available and more open to unexpected discoveries. 

As noted by particle physicists Karagiorgi et al.:
``The development and deployment of machine learning methods in the search for new fundamental physics is becoming urgent given the dearth of evidence from traditional methods'' \citep[p.~10]{karagiorgi2021machine}.
The traditional methods are top-down, model-dependent searches from particular BSM models.
The recent failure of this approach pushes physicists towards more model-independent, bottom-up methods that make fewer assumptions about the target models of the searches, encouraging physicists to remain open for unexpected discovery.\footnote{Of course, many physicists are still working on high energy theories like supersymmetry and model-based searches generally; model independence is a fast growing part of a many-pronged, but overall more open-ended search strategy.} 
Performing a model-independent search, such as using an autoencoder network for anomaly detection, has a huge benefit that one does not need to devise a new physics model, perform calculations and simulations and arrive at a precise signal that one could see in a particle detector, compare this signal to the real data and see if it statistically likely to be there. 
Rather, one aims to capture deviations from known physics as well as possible, without specifying what those deviations will look like.


The focus of this paper will be on the model independence of DL approaches, where analyses are not necessarily based on physicist-identified features, and in particular, on unsupervised learning methods, where the training of the network is more hands-off. 
Applying these DL models to find new physics is quite involved and there are many issues about choosing and designing the model architecture, how to train and verify the algorithms, how to structure the loss function, and how one is to interpret the outputs that still must be handled by physicists (more on this below).
However, the hope is that they may cover spots blind spots in model-based approaches, together forming a more complete search strategy.

\subsection{Spectrum of Model Independence}		\label{sub:mi}

So what, more precisely is model independence? 
No model is completely independent of models and no DL training is completely unsupervised, but that does not mean that model independence is not a useful and important concept. 
As a first working hypothesis, one can characterize model independence as the aim to reduce model (or theory) bias. 
This is a contrastive definition (from the nature of the word `reduce'), but it seems the best way to characterize it: something is model independent if it is less model-dependent.
This reflects that strategies and methods may be more or less model independent than others; it is a property that comes in degrees. 
There are a great number of modelling assumptions that may go into a search, calculation, or measurement and these can be lessened to various degrees. 
Model-independence is thus a broad term for a family of strategies that can be pursued for many distinct reasons. 

Let us begin at one end of the spectrum of strategies for finding new physics. 
There are model-based methods, which may involve deriving predictions from a given model and then attempting to experimentally verify them. 
Searches can exploit detailed features of a given model to constrain the parameters of the hypothesis being tested. 
One has to restrict possible measurements to those where the model is expected to contribute significantly, or to exploit correlations between different measurements.
This makes for more efficient analyses, but is only narrowly applicable to cases where the modelling assumptions are satisfied. 
This can be a worthwhile strategy when one has good reasons to employ the assumptions of that model or restrict the values of certain parameters, etc.
Such a search in HEP might be to look for the signal of a charged Higgs boson in a type-II 2-Higgs doublet model with minimal supersymmetry. 
One will get a very good idea what this signal should look like, which will help in determining if the signal is there, but it will have little to nothing to say about even a slightly different model. 

There are also moderately model-independent methods, which one could call `model-agnostic'.
These methods relax the dependence on any one particular model, but still depend significantly on some modelling assumptions. 
One popular example is by using \textit{simplified models} (see \citep{bechtleetal} and \citep{mccoymassimi} for a philosophical discussion of model independence and simplified models). 
By using a simplified model, one can perform a search for a given particle, say a stop quark, that could be a part of many different supersymmetric models, without specifying the rest of the target model.  
One can greatly simplify the search by assuming that that the particles other than the stop quark are sufficiently heavy enough to have little effect on the target signal. 
Results of the search may be applicable to a whole class of new physics scenarios where these assumptions apply.
It should be noted that this lack of dependence on one particular model is sometimes referred to as `model independent' (e.g. by \citet{mccoymassimi}). 
However, the term `model agnostic' is more apt since there are considerable modelling assumptions that go into calculating the target signal.
It is a signature-based search that is applicable to even non-supersymmetric target models, but it is aimed at target models (or target particles), which makes it distinct from what I am calling model-independent searches.
This agnosticism is an important feature of some searches since it can be much more efficient and can indicate that the results are unlikely to be the artefacts of a particular model. 
These search methods aim at reducing modelling assumptions and biases, but the specificity of the target means that one is unlikely to find something that one is not already looking for. 
By contrast, some physicists have used the term `model agnostic' to describe methods that I count as model independent \citep[for example,][]{Dillon:2023zac,Matos:2024ggs}. 
They often cite the reason that nothing is completely independent of models, but this is not a reason to avoid the term, because the role of a model is here to define the background. 

In order to bring the use of terminology into line, we must make a distinction between the \textit{target model} and the \textit{background model}. 
Such a distinction has already been put into practice in handling problems of circularity and theory-ladenness.\footnote{A similar distinction has been drawn in many places in the literature, for example, between theory in model construction and in model use in \citep{morrison99} and the necessary but harmless role theory plays was also discussed in \citep{Franklin2005-FRAEE}.}
When one conducts top-down model testing, one is making assumptions about the target of the search, but when one conducts bottom-up searches for deviations, one only needs to assume features of the background against which something different can be detected.
Hence, the important feature of model agnosticism is to be open to several or many target models, and the important feature of model independence is to reduce dependence on any target models whatsoever. 

Towards the model-independent end of the spectrum we have searches that strongly reduce the assumptions and biases of target models, such as by performing precision measurements on SM observables. 
Precision measurements can provide a window to BSM physics, since many values of SM parameters are precisely determined and constrained by theory and by the values of other observables. 
Deviations are most likely to be due to the effects of new physics.
A philosophical take on such precision measurements was recently undertaken by \citep{Koberinski2020-KOBQQ} looking at the case of the muon $g-2$ anomaly.
Precision measurements can be aided by a parametrization scheme like the Standard Model Effective Field Theory (SMEFT).
In a recent paper, \citet{bechtleetal} undertake a detailed case study of the SMEFT and the bottom-up approach to finding new physics.
The idea behind this approach is to expand the SM with all possible effective operators for new interactions, but not to add any new fields. 
One then uses this as a scorekeeping device to parametrize deviations from the SM. 
The coefficients on these new operators are tuned with data and if the best fit is non-zero, then there is evidence that the operator or set of operators might be relevant in some new physics processes. 
This method does not make use of predictions from specific BSM models and opens up the possibilities for unexpected discoveries.

Here at this end of the spectrum we also find DL-based searches, such as those done with autoencoders.
The very basic idea is that a deep neural network (DNN) can be trained to learn the SM from simulated data. 
This is then compared to real data and an anomaly score can be calculated to see if there are differences from what one would expect given the SM alone that might indicate new physics. 
Thus, only a well-defined background model is necessary and the role of a target model is reduced to benchmarking, optimization, or later-stage interpretation. 
We turn to this in the following section. 

\section{Deep Learning} \label{sec:dl}

DNNs are networks for decision making that consist of many layers of interconnected nodes that are thought to resemble the arrangement of neurons in the brain (hence the name).
Each node computes an \textit{activation function} from its input and passes on its output signal along links to the following layer from the input layer to the final output layer, which is then interpreted as the network's decision for that input.
The strengths of the links between the layers of nodes are parametrized by values called \textit{weights} that amplify or dampen the signals transmitted through the network's layers. 
The learning process involves iteratively updating the weights to minimize the difference between the network's decisions and the desired outputs, which is calculated with a \textit{loss function}, like a mean-squared error. 
The weights are often initially randomized and are adjusted via a process called backpropagation, which uses the chain rule to compute how weights affect the loss function.
The network architecture is set up and hyperparameters, such as the number of layers, are decided, but the network is not programmed in the traditional sense. 
The network is trained to achieve a certain goal like classifying images correctly, but how it does so is not programmed, but a product of the training. 
A computer program can also solve similar problems, but it will do so in the way that it has been programmed to.
A DNN by contrast finds its own way of solving the problems posed to it. 

The main practical goal of machine learning is generalization.
The models are trained to learn some set of data, but are intended to perform similar decision tasks on further data. 
The data must be learned, but what is learned must be generalizable. 
There are many methods to accomplish this. 
Typically, the data is divided into a training set and a test set.
The model is trained on a portion of the data (something like 75\% of the data) and then the model's predictions are tested against the remaining (25\% of the) data that it has not seen before. 
Further supervision can be imposed by hand in the sense that trainers can explicitly tell the network when its outputs are not as desired.
Other methods of regularization to avoid over-fitting are to vary the training data in various ways, to add penalty terms, or using `drop out' to turn off certain nodes in the model (see \citep{kukacka2017} for an overview).

We can distinguish two distinct qualities of well-trained, well-performing models: \textit{reliability} and \textit{robustness}\footnote{These terms are the subject of considerable philosophical debate, and I do not wish to survey the various uses and contribute in any way to this debate. I will simply use them as I describe them here. The terms are here used along similar lines as they are described in \citep{grotesullivan}, which one can see for a more detailed discussion.}.
The reliability of a model is an indication of the model's ability to produce similar outputs for similar inputs. 
A model is unreliable if the same input, or relevantly similar inputs, produce drastically different and in at least some cases, undesirable outputs. 
Reliability is essentially an indication that the model has learned relevant features of the training data; that it is not underfit. 
As mentioned, one also wants to be sure that the model is not overfit and for this one can test for robustness.
A robust model will be generalizable to new data beyond the training and test data.
The goal of training is to have models that are both reliable and robust and as we will see, ensuring this will go a long way towards alleviating concerns brought up in Sec.~\ref{subsub:risks}.

\subsection{Autoencoders}		\label{sub:dl_basic}

AENs are deep convolutional neural networks that work by taking an input and transforming it to a low-dimensional representation in latent space then elaborating it back to a high-dimensional representation to resemble the input as closely as possible.
They are trained to minimize an error that is calculated as the difference between the output and the input. 
They are widely used for de-noising, data compression, and feature extraction due to how efficiently they form low-dimensional latent representations. 

\begin{figure}[ht] 
	\includegraphics[width=\textwidth]{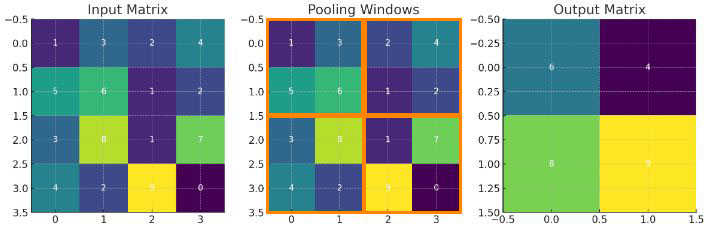}
	\caption{Example of max pooling where we see the original 4x4 matrix on the left, the 2x2 windows in the center, and the extracted 2x2 matrix on the right featuring only the maximum values. \label{fig:maxpool}}
\end{figure}

They consist of an encoder and a decoder network that are similar but not always identical. 
The encoder leverages convolution to deal with the high-dimensional input space.
Convolution is a linear operation that passes a kernel (essentially a bank of filters) over the data to amplify desired features.
Typically, a convolutional node often passes its output to a \textbf{ReLU} (rectified linear unit) which activates according to whether its input is above a certain threshold (sometimes called the \textit{detector stage}). 
For example, a kernel that amplified vertical lines and passed this on to ReLU unit would then return an image with only vertical lines. 
In order to return lines at any orientation, one needs to pass the output through a third node that aggregates and `downsamples' the activity of several different filters with overlapping spatial receptivity. 
The most common downsampling function is \textbf{max pooling}, which is the process of pooling only the maximum values for a given field (a subsection of the image) (Fig.~\ref{fig:maxpool}).
The network performs a kind of abstraction, no longer having access only to pixels, but to high-level features in a latent space representation. 

In a second set of steps, the decoder then proceeds to upsample the latent representation, filling back in a larger matrix by various means, such as max unpooling, nearest neighbor, or transpose convolution methods, e.g. 
Because the output is generated only from the salient features captured in the latent representation, the output will be relatively free of the high spatial frequency noise of the original input.
If the input is too noisy then the network will capture and enhance parts of the noise rather than the intended signal and will lead to a high reconstruction error. 
A well-trained network will have high reconstruction errors only for inputs sufficiently different from its training data.
This can be insufficient for anomaly detection if the signal is too similar to the background or too easy for the network to reconstruct. 

In order to calculate the error, or loss, one needs a metric to calculate the distance between events---between the input and the reconstructed data. 
This can be accomplished in various ways, such as by mean average error, mean squared error, or optimal transport distances. 
Further, a threshold above which one considers the difference to be an anomaly needs to be established, e.g. by some chosen number of standard deviations. 

Variational autoencoders (VAEs) are a special kind of AEN where input data is mapped to a distribution rather than to a point in the latent space. 
The latent space generated is a (typically Gaussian) distribution that is then sampled back to the original space. 
This makes VAEs especially useful for particle physics as it can encode the probability distribution of the SM background training sample. 
The VAE essentially calculates how likely the input is on the assumption that it came from the training data. 
This gives an anomaly score based on the difference between the probability distributions plus a Kullback-Leibler divergence term that regularizes the latent space and allows for meaningful sampling. 

\subsection{Deep Learning in HEP}	\label{sub:dl_hep}

Despite being at the early stages of DL in HEP, there have already been significant demonstrations improved performance. 
For example, various kinds of DNNs have been used in many different tasks and shown to have comparable or even significantly improved performance over traditional methods in distinguishing dark matter signatures in LHC physics \citep{Khosa:2019kxd}, in searching for exotic Higgs decays \citep{Jung:2021tym}, and in jet flavor tagging \citep{Munoz:2022gjq}, top tagging \citep{kasieczka_mltop}, anomalous jet tagging \citep{Cheng:2023dal}, optimizing the reduction of a nuisance parameter \citep{agnolo2019}, to improve the triggers at the LHC \citep{Pol:2020weg}, and many more examples are emerging every day. 
In this section, I will describe a few ways in which DL is used in HEP with a focus on how unsupervised methods (and less-than-supervised methods generally) can help search for new physics in a model-independent manner. 

Machine Learning has been around in HEP for decades\footnote{See \citep{albertsson2019machine}, \citep{Bourilkov:2019yoi}, \citep{Duarte:2024lsg}, or \citep{guestcranmer} for an overview of ML in HEP.} in the form of boosted decision trees (BDTs) and multivariate analysis, but the recent advances in deep learning have changed the game. 
As early as 2014, it has been remarked that deep learning could be a very promising analysis tool: ``Our analysis shows that recent advances in deep learning techniques may lift these limitations by automatically discovering powerful non-linear feature combinations and providing better discrimination power than current classifiers---even when aided by manually-constructed features,''\citep[p.~9]{baldiwhiteson}.
More recently, Schwartz has stressed the dramatic changes that have come with the shift from `traditional' shallow ML to `modern' DL:
\begin{quote}
	``In the relatively few years that modern machine learning has existed, it has already made traditional collider physics obsolete. In the past, physicists, including me, would devote their efforts to understanding signatures of particular particles or processes from first-principles: why should a stream of pions coming from a W boson decay look different than a stream coming from an energetic gluon? Now we simply simulate the events, and let neural network learn to tell the two samples apart.''\footnote{The first sentence is surely exaggerated, but what Schwartz probably means is that modern methods (i.e. deep learning) has been shown to outperform shallow machine learning methods.}
\end{quote}
\begin{flushright}
	\citep[p.~10]{schwartz2021modern}
\end{flushright}
What Schwartz remarks is that there is a reduced role for physics knowledge---physicist input---in the analysis.
And yet, surprisingly, this machine-learning approach often outperforms traditional methods.
Traditional analyses are performed with respect to a single variable and multivariate analysis marked a drastic improvement.
But today, state of the art machine learning has come a long way from mere multivariate analysis and moves into what can be called \textit{hypervariate} analysis, where ``the entire phase space of available experimental information can analysed holistically'' \citep[p.~1]{karagiorgi2021machine}.
Traditionally, in searches for new physics the data is compared to simulated outputs of what one might expect to see in the detector given various modelling parameters---the data are optimized and interpreted in the context of specific models.
It is not possible to compare all the data to all possible models in this way, but DL methods may provide a strategy that is ``liberated from model dependence compared with traditional searches''\citep[p.~2]{karagiorgi2021machine}. 

Most ML tasks in HEP can be formulated in terms of the optimization of a loss function $L[y,f(x)]$.
The aim is to search for the function $f(x)$ that optimizes move from high-dimensional space of observed data to a low-dimensional space. 
The (reduced) space of functions searched is given by a series of transformations mapping inputs $x$ onto hidden states $h_i$ and then to the output $y$, $h_{i+1}=g_i(W_ih_i+b_i)$, where $g_i$ is the activation function and a particular $h_i$ is the $i$th transformation, called the embedding, and the $W$ weights, and the $b$s are biases.
Training examples are used to calculate the gradient of the loss function with respect to the model parameters $\nabla_\phi L[y, f_\phi(x)]$.

The tasks of HEP are often several steps removed from searching for a new particle.
DL is used at just about every stage and it has many different tasks---it is not always performing model selection for discoveries, but may be used to make a binary decision whether a given particle was, for example, a top quark or not or to generate probability distributions.
It can also be used, of course, for determining whether a measured probability distribution was likely to have been generated by a given BSM model. 
It is important to keep in mind that there are many different uses for DNNs in particle physics with different tasks and architectures. 
Just as with models, ML networks can also be placed on a spectrum of model independence. 

There are \textit{supervised} learning methods, which are motivated by particular signals. 
These are relatively model-dependent, given that the search is optimized for a target model or a given signature (either object tagging or event classification). 
Here, simulations of the SM and simulations of some new particles are used and classifiers then separate potential signals from the background in real data. 
Networks can also be \textit{less-than-supervised}, which do not have the signal hypothesis as part of the test, but merely attempt to identify background and non-background. 
These come in two primarily sorts: \textit{weakly-} and \textit{semi-supervised} learning methods\footnote{Weakly supervised methods use noisy data, and semi-supervised methods use noiseless, but partial data. The methods are similar other than the fidelity of the labels.} used to estimate a ratio of the proportion of two data sets (labelled signal-like and background-like) in a third test set (real data). 
Lastly, there are \textit{unsupervised} methods that use unlabeled data and employ, e.g. AENs. 
These are called unsupervised because there is no need to specify a special or desired target signal---the network merely learns to represent the background on its own (e.g., in latent space) and later on can be used to flag anomalous data.
Sometimes distinctions are made as to whether the network creates labels or merely clusters data or estimates density (the latter being referred to as \textit{self-supervised}).
Modern DL often blurs and mixes different so-called `learning paradigms' and so keeping them neatly separated is a bit of an academic exercise. 
What is important is that through DL, networks can find their own solutions to problems so that they may stumble upon new and more optimal solutions.
DNNs, such as AENs, have the potential to be strongly model-independent and so we turn to these now.

\subsection{Hunting for Anomalies} 		\label{sub:hunting}

The basic programme for anomaly detection is to first select and design a network and set various hyperparameters, such as number, arrangement, and sizes of layers, including the bottleneck layer size for AENs, which importantly determines the latent representation.
The second step is to preprocess the data to be suitable for the network. 
This can involve rotating and centering images, choosing the pixel size of the images, and de-noising.
The third step is to train and optimize the network and have it learn to reconstruct the SM background with a small reconstruction error.
This is often done from simulated datasets, so that there are true labels for events, but this is not necessary.\footnote{There is a popular framework for working without labels, called CWoLa (Classification Without Labels), and one can use self-supervised methods that create and employ `pseudo labels'.} 
The fourth step is to do BSM benchmarking and ensure that the model and the loss function give large reconstruction errors for a variety of BSM scenarios, for additional $W'$, $Z'$, leptoquarks, charged Higgs, etc. and set a threshold for anomaly that captures many BSM signals but does not capture SM processes or noise. 
The final step is to test for anomalies in the data and to study any flagged regions with various other methods and resources.

There are two basic methods to searching for anomalies \citep{Fraser_2022}. 
For the first, one tests if two samples, one from drawn from SM distributions and the other from real data, are likely to have been drawn from the same probability density distribution. 
This is done where signal events are likely to look similar to the background. 
In such cases, a network can learn the background distribution, subtract this from the real data, which is essentially signal+background (if there is a signal), leaving one with a clearer indication of a new signal.
Here one essentially uses the network to clean the background, leveraging autoencoders' de-noising abilities, then performing a separate bump hunting search for anomalies. 
The second method is to search for anomalous individual events directly. 
After learning a distribution, each event can be given an anomaly score, given naturally as the distance between an event and the learned distribution (or a fiducial event representing the learned distribution).

The hunt for anomalies has become such an important approach that it was recently the topic of two major competitions, the LHC Olympics \citep{Kasieczka_2021} and Dark Machines \citep{darkmachines}.\footnote{CERNs largest experiments, CMS and ATLAS, also recently reported much success with novel anomaly detection with DL at the International Conference on High Energy Physics (ICHEP 2024) \citep{Kaur:2915241,D'Avanzo:2910974}} 
The LHC Olympics was an open community challenge to explore various approaches to anomaly detection with machine learning.
The aim is to test the networks on three blinded datasets possibly containing an anomaly.\footnote{There were 18 individual approaches that were submitted, about half of which used unsupervised methods, the other half were weakly- and semi-supervised. For more details on the setup of the competition and the individual methods, see \citep{Kasieczka_2021}.} 
On Black Box 1, about half of the approaches were able to identify the correct resonance within a reasonable window.
Black Box 2 had no signal injected, but many approaches predicted resonances. 
Black Box 3 had a signal with two decay modes, but was not correctly predicted by the models and only by some after unblinding.
None of the methods was strictly better than all the others, each demonstrating different capabilities and different optimal applications and vulnerabilities. 
The more strongly supervised networks performed well when the signal was known (i.e. after unblinding), but more poorly on the blinded datasets and when the signal was different than expected.
This highlights the sensitivity of performance to the network's architecture and to the kind of signal that might ultimately be discovered.
If one's assumptions about the features of the new physics signal are correct, then supervised models to discover these can be successful. 
Otherwise, unsupervised methods seem to perform better, but the authors stress that ``there still is substantial room for improvement for anomaly detection in the realm of particle physics'' \citep[p.~90]{Kasieczka_2021}.

The Dark Machines initiative held a competition for anomaly scores as part of the group on ``unsupervised searches at colliders'' \citep{darkmachines}.
The competition involved two generated datasets based on SM background simulations injected with different new physics scenarios and over one thousand algorithms, many based on variational autoencoders as well as flow-based likelihoods.
The major objective was to find the algorithms that worked best on the most new physics scenarios. 
They calculated the `best' in a number of different ways, and the details of given approaches are not so important, but I would like to stress an important takeaway.
Most of the best algorithms still had some signals with essentially no improvement and some algorithms with strong improvements in some signals had low improvements in others.
This indicates that the success of anomaly detection is dependent on the particular new physics scenario.
Nonetheless, many of the best algorithms demonstrated strong improvements in significance of new physics signals in many channels.\footnote{Significance Improvement is a measure of the change of signal events over background events after the application of the anomaly detection. They note that ``This metric does not tell us if the anomaly detection technique is capable of discovering new physics, as this still depends on the cross sections, but it informs us on how much the anomaly detector can enhance the statistical purity of the signal over the SM noise'' \citep[p.~40]{darkmachines}.}

\begin{figure}[!ht]	
	\includegraphics[width=\textwidth]{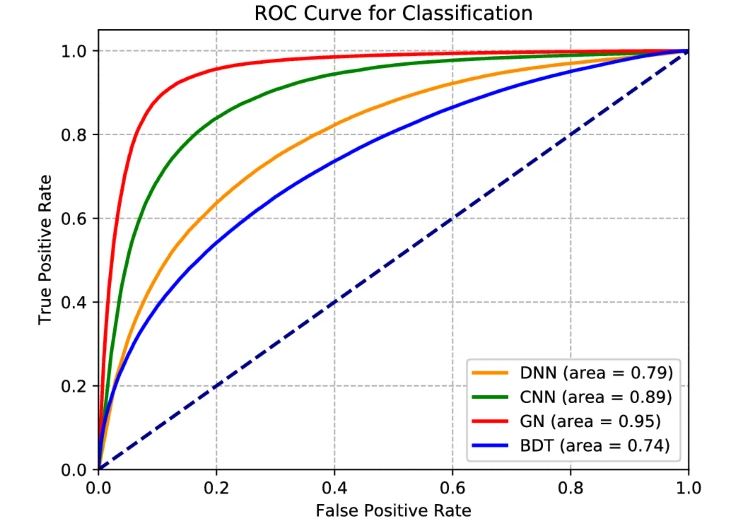}
	\caption{Three different deep learning classifiers, labelled \textit{DNN}, \textit{CNN}, and \textit{(GN) GoogleNet} compared with the traditional BDT in the task of distinguishing $\gamma$ vs. $\pi^0$. These deep networks featured between three and fifteen million trainable parameters. For more benchmarking performances, see \citep{Belayneh_2020}.}
	\label{fig:benchmark}
\end{figure}

Competitions like this are how physicists are learning which architectures and algorithms are most sensitive to a broader range of new physics scenarios and which are less efficient and have more limited applications.   
Improvements in this domain will be made through trial and error. 
It is important to note that in the world of DL, these approaches used very small networks with a few layers and a few hundred to a few thousand tunable parameters. 
Other approaches that have presented impressive benchmark figures are leveraging enormous networks with tens of millions of parameters (this is to say nothing of large LLMs that feature trillions of parameters). 
One such study was presented in \citep{Belayneh_2020} to distinguish a photon from a pion (Fig.~\ref{fig:benchmark}). 
They used three different deep learning classifiers and one traditional state-of-the-art classifier, the boosted decision tree. 
Fig.~\ref{fig:benchmark} plots the true positive with the false positive rate for each and the area under the curve essentially signifies the classifier's performance, maximizing the true positives and minimizing the false positives.
All the deep networks outperformed the traditional ML approach in this task.
This is not anomaly detection, but highlights a different successful application of DL in HEP that is aided by the size of the networks and may translate to improved success in anomaly detection in the future.

\begin{figure}[!ht]	
	\includegraphics[width=\textwidth]{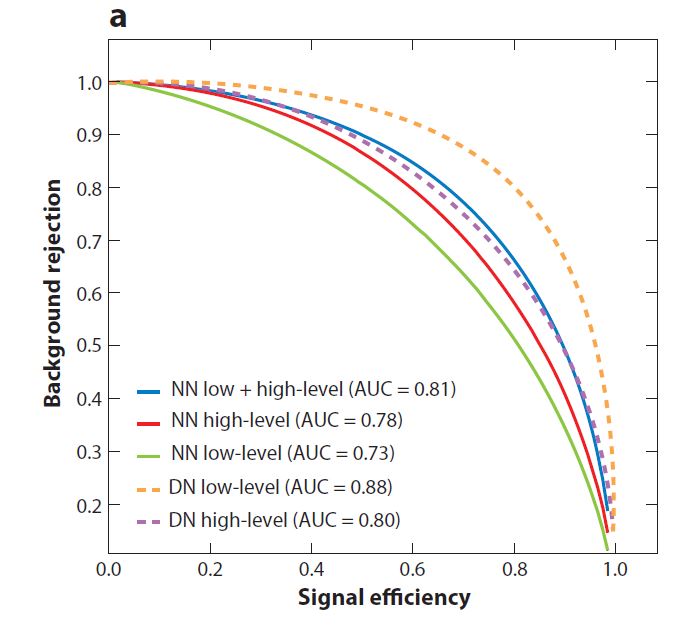}
	\caption{This plot compares different networks in background rejection and signal efficiency. Both are desirable, so the Area Under the Curve (AUC) shows the network's performance.}
	\label{fig:outperform}
\end{figure}

An important task in hunting for anomalies is to identify a signal above a background. 
What one wants to know for the prospects of DL is whether the network can more efficiently learn the relevant features of the data itself from low-level data (like four vectors or image pixels) or whether the best performance comes when the network is told what high-level features to look for.
In a review article covering many promising machine learning studies, \citep{guestcranmer} look at signal-to-background classification comparing deep and shallow neural networks (Fig.~\ref{fig:outperform}). 
In their own words, ``A comparison between the performance of deep networks (DNs) in signal–background classification and that of shallow networks (NNs) with a variety of low- and high-level features demonstrates that DNs with only low-level features outperform all other approaches,'' \citep[p.~169]{guestcranmer}.

This is echoed by \citep{Baldi:2022okj}, who look at separating signal from background in exotic Higgs decays. 
They note that before DL techniques, one would often train shallow networks and BDTs for classification with features built using expert knowledge, which would boost performance. 
In their benchmark tests ``The shallow neural networks and BDTs trained with the high-level features perform significantly better than those trained on only the low-level features, demonstrating the importance of feature engineering in shallow
machine learning models\ldots only the deep learning approach shows nearly equal performance using the low-level features and the complete features. This suggests that it is automatically discovering high-level abstractions similar to those captured by the hand-engineered features, obviating the need for laborious feature engineering,'' \citep[p.~6--7]{Baldi:2022okj}. 
This is also noted by \citep[p.~5]{jawahar_vae} that present their results from the Dark Machines competition.
Their results show that``adding high-level features brings no definitive improvement in performance, thereby leading us to conclude that the baseline model with marginally lower number of trainable parameters is a good choice.''

\citep{farina2020} use an AEN as an anomaly detector. 
They look at QCD jet images as background and boosted top jets and r-violating gluino jets as a signal. 
Since AENs are well-suited to noise reduction, they opt to clean the (QCD) background with the autoencoder, then hunt for bumps in jet mass. 
They show that the signal-to-background ratio with the autoencoder improves almost 6-fold over the search without.
``With an autoencoder approach, one does not need to know what one is looking for. It is a powerful new method to search for any signal of new physics in the data, without prejudice,'' \citep[p.~10--11]{farina2020}.

\citet{Andrews_2020} have proposed an end-to-end ML strategy for particle physics based on image classifiers.
The method is to employ a convolutional DNN on low-level data and perform high, event-level classification. 
The networks are capable of learning the angular distributions and shapes of electromagnetic showers, even though the individual particles are not resolved as an intermediary step.
The network operates on tracks and calorimeter hits.
The detector (CMS) is essentially a 2D projectable surface and the data can be easily formatted as a rectangular image. 
They show that using this method they can distinguish SM Higgs decays from its background sources.
The hope is that end-to-end DL could drastically improve the model-independence of the search, but it has yet to be fully implemented.


\subsubsection{Epistemic Issues}		\label{subsub:risks}

While the approach brings breadth and many improvements over traditional searches, there are difficulties involved and there are compromises to be made with respect to the model independence that can be achieved---it is not a pure inductivism. 
I will here review three kinds of issues as they pertain to model independence: 1) technical issues, such as choices of network architecture; 2) the heavy use of simulation; and 3) the interpretability of the outputs. 
There are other philosophical worries that one may have about research with deep learning in general, such as the lack of mathematical theory and the opacity of the networks, but these are less relevant for considerations of model independence where the target models of new physics are at issue.

Let us begin with the more technical issues. 
As DNNs are essentially mathematical transformations, much of the question of their model independence hinges on their input. 
While weights and biases can be tuned by the data, many modelling choices must be made prior to use. 
These choices remain in at least the following areas:

\begin{enumerate}[i]
	\item \textbf{Learning Paradigm.} Physicists need to choose what kind of learning method to use and decide whether supervised, or semi-supervised, unsupervised learning or some hybrid approach is most appropriate. If one is performing a search for anomalies, perhaps unsupervised or weakly-supervised learning will be the most effective, but to tell two known samples apart, a fully supervised network will likely be the most sensitive. Given known results, it is already pretty clear given the task what approach will work best.
	\item \textbf{Hyperparameters.} There are choices in the hyperparameters, design, and architecture of the networks: how big should the network be? Should it have convolutional layers? Should it be an autoencoder? Even if one has decided this there are further optimization questions, such as how many downsampling and upsampling blocks should there be? What should be the size of the latent space? What is the anomaly threshold? etc. However, studies can often optimize their own networks simply by trying out different arrangements and taking what works best, comparing 30 node dense layers with 100 node layers and 40x40 pixel input images with 64x64, and so on. 
	\item \textbf{Loss Function.} The network's main task is often to minimize a loss function, but this loss function needs to be chosen. There are different tasks that may benefit from a different kind of loss function and so appropriate ones need to be tried and selected. For autoencoders, one should consider how the errors will be calculated. \citep{Fraser_2022} performed their analysis comparing mean squared error, mean average error, and $p$-Wasserstein distance metrics to see which give the best AUC and signal efficiencies for distinguishing top and $W$ jets from QCD jets. Again, many studies will compare the results of different choices. 
	\item \textbf{Data Preprocessing.} Data need to be preprocessed. Jet images for example need to be flipped, rotated, boosted, pixelated, and so on. These processes must be neutral with respect to the results. In general, training data samples must be iid, there cannot be too much admixture of signal in the training background, and the simulations must be reliable.
	\item \textbf{Signal Assumptions.} There are assumptions about the nature of the anomaly sought. These assumption creeps up in many of the other decisions physicists have to make (i--iv), and lies at the heart of the second and third kinds of issues (simulation and interpretation). One optimizes a network and the loss function by testing on given signal identification tasks. In general one makes sure an AEN works properly by making sure it flags particular BSM scenarios as anomalous. One can therefore only have confirmation that the network works on some set of known BSM scenarios. There is no guarantee that a BSM signal will not also be faithfully reconstructed by the AEN, if it is for whatever reason `easy' to reconstruct or if the new physics simply does not involve a sharp resonant peak above the background. In the LHC Olympics for example, the focus was on resonant signals, but in a non-resonant case, there is no general method of estimating the background \citep[p.~92]{Kasieczka_2021}. One need not use BSM signals, but searches utilizing just the background are not as effective as those that are also trained on potential BSM signals \citep[see, e.g.][]{Matos:2024ggs}. Training on a variety of BSM scenarios gives credence that the network is not dependent on any given signal to flag an anomaly, but one could harbor reservations that the AEN is more model agnostic than model-independent. 
\end{enumerate}

This brings us to the second kind of issue, namely, the use of simulation.
The training of DL models in HEP is done mostly, or in some cases exclusively, with simulated data. 
This leads to two distinct worries: first that this negatively affects arguments about its generalizability, that is, its trustworthy application to real data; second, that modelling assumptions are smuggled into the network from those of the simulated training data.
In general, ML models are biased from two sources: they inherit biases from the data (in the same way that language models learn from existing bodies of text that contain biases) and biases are explicitly built in by modelers (as listed above).

The challenge to claims of model independence comes from worries of strong theory-laden assumptions in the training data. 
When it comes to theory-ladenness in the training data for the SM background, this is of course no problem---the very idea is that the network should learn the background theory as well as possible. 
However, as mentioned with point (v) using a network that has been  tested on certain BSM scenarios rather than others and deciding how one chooses the best network after testing reveals assumptions about the nature of the signal being sought. 

The fact that the models are heavily reliant on simulation is not itself a problem. 
Networks are also often trained on real data, a mix of real and simulated data, or are tested against real data. 
One can train on real data, but there is no true labelling to rely on. 
One does not know how a given event datum was produced, and cannot know, even in principle, since the decays are quantum processes.\footnote{There are of course reliable statistics, but this does not equate to a true label for a given event.} 
As mentioned in Sec.~\ref{sub:hunting}, there are methods to work without true labels that have demonstrated great success. 
For example, on can find `clean' events or sideband areas and train on this rather unambiguously \citep{PhysRevLett.120.241602}. 
This is done in $t\bar{t}$ production, where the anti-top decays into a muon and a $b$-quark jet that can be clearly tagged. 
The rest of the information can then be cleanly analysed as top quark data. 
This is called the \textit{tag and probe} method and is a pillar of HEP \citep{schwartz2021modern}.

Simulations in HEP are validated by material measurement where possible.
\citet{Mattig2021-MTTTSA} outlines under what conditions simulations become trustworthy. 
However, he notes that contrary to other claims, such as by \citet{morrison2015}, their credibility is not on par with material measurement and their trustworthiness requires, among things, significant agreement ``with many material measurements in a large region of property space'' \citep[p.~14440]{Mattig2021-MTTTSA}.
Experimental measurements must epistemically secure the results of simulations in order for those simulations to be trustworthy.
Where simulations can take the place of real measurements, they do so only insofar as they have been rigorously validated. 
One way to verify that a simulation is accurate is to learn where it is inaccurate.
One can do so by having a generative network learn to make its output indistinguishable from the real data. 
This iterative re-weighting of simulated datasets is performed by Omnifold \citep{PhysRevLett.124.182001}, for example, to try to discover and remove the artefacts of detector simulation, a process called \textit{unfolding} \citep{schwartz2021modern}.
Simulations of known physics are in general very well handled.

This brings us to our final kind of issue, namely, the assumptions about new physics that are required to get meaningful interpretations out of the network's outputs. 
The teams in the LHC Olympics did always not succeed in detecting the anomalous signal from the lowest-level data: ``None of the methods deployed for the LHC Olympics were able to find anomalies using the full list of hadron four-vectors directly---successful approaches all used some model-inspired dimensionality reduction. Scaling up these approaches to high dimensional feature spaces is a key challenge for the next years and will require both methodological and computational innovation,'' \citep[p.~92]{Kasieczka_2021}. 
The data had to be processed into a lower-dimensional feature space in order for the approaches to succeed, but there are reasons to be optimistic. 
In the benchmarking tests shown above (Figures~\ref{fig:benchmark} and \ref{fig:outperform}), very large networks transformed high-dimensional data into a lower-dimensional latent space, without physicists programming in what those features are.
In ordinary computer vision tasks, unsupervised convolutional networks learn to identify low-dimensional features like eyes, hands, and car tires out of the high-dimensional pixels of the training data images. 
As mentioned in Sec.~\ref{sub:hunting}, the LHC Olympics networks were comparatively quite small and in the few years since, much progress has been made in these dimensional-reduction tasks such that they are among the most successful tasks of DL (see, e.g. \citep{VELLIANGIRI2019104} for an overview of DL dimensional reduction techniques).

Interpretability is a particular issue in deep learning because the correlations discovered by large networks are not naturally presentable in terms of histograms and contour plots comparing quantities that physicists are used to. 
The power of the networks to discover highly non-linear combinations among a huge number of parameters is not necessarily easy to exploit. 
There are of course efforts to take what are called `explainable AI' approaches to HEP (see \citep{neubauer2022explainable} for an overview). 
Explainable AI, or interpretable ML, is still a new field in HEP and I will only mention two kinds of approaches that are being attempted: i) working directly with interpretable models and ii) modelling DNNs with shallow models.
The first approach was shown to be successful in \citep{Lai:2020byl}, where a constrained GAN was able to learn the final distribution of particles in a gluon shower and learned the underlying mechanism in a way that could be understood by a human physicist. 
Typically, one sacrifices performance of deep networks for the interpretability of results, but improvements can be made to networks where the connections between the input, output, and network architecture are clearer. 
The JUNIPR framework \citep{Andreassen_2019} is also worth mentioning here. 
It is one of several attempts to employ unsupervised learning with a an interpretable network architecture.
The outputs are given in terms of probabilities attached to jet clustering trees. 
Their approach ``forces the machine to speak a language familiar to physicists, thus enabling its users to interpret the underlying physics it has learned'' \citep[p.~4]{Andreassen_2019}.
The second kind of approach is to attempt to provide a post hoc explanation of a DNN's performance by mimicking it with a shallow network; a strategy called \textit{model compression} (see \citep{cheng2020survey} for a survey on DL model compression).
One way to work around the better performance of DNNs is to train the shallow network on so-called `soft' labels produced by the deep network, rather than on the real data; effectively using the DNN as a teacher.
\citep{francescato2021} used this method to build light, simple architectures with minimal performance loss from the DNN in a simulation of a planned trigger for the high-luminosity LHC.

The difficulties with interpretation come in at the later stages of the anomaly detection program outlined in the beginning of Sec.~\ref{sub:hunting}. 
Even if an anomaly can be detected, it does not mean that this is a signal of BSM physics and it may give little indication of what new physics lies behind the deviation.
As mentioned, the model-independent approach itself will not provide evidence for a new physics model, but only signal where traditional model building and model testing takes over, though perhaps aided once again by other DL approaches.
A 5-sigma result of a statistically significant deviation from the SM will not tell physicists what the new physics is, but it will tell them where to look. 
This is rather similar to the successful application of the SMEFT framework \citep[see][for details on this process]{bechtleetal} or to precision measurements in general. 
Thus, even though the DL network does not provide an interpretable model, it may greatly aid in the discovery of new physics.



\section{Model Independence Revisited}	\label{sec:lessons}

The aim of model-independent methods is to take some key choices out of the hands of physicists, relaxing target model assumptions, and letting the data guide the way.  
I had preliminarily characterized model independence as having the following characteristic aim: it significantly reduces modelling assumptions and biases.
We can now more precisely specify two necessary conditions.
A search method is model-independent if it:
\begin{enumerate}
	\item has no target model; and
	\item has a well-defined background against which deviations can be seen.
\end{enumerate}

This is most clearly satisfied by something like a precision measurement of a SM observable. 
Theory and modelling is still present here, but there is no target model of BSM physics---the target of the measurement is the background. 
In the model-independent searches using DNNs, there is similarly no target model. 
Assumptions about likely BSM scenarios enter in the benchmarking in the anomaly testing as described above, but the network remains open to deviations by new physics of almost any kind (though given the inevitable specificity of the benchmarking, there are no guarantees of high anomaly scores for unexpected deviations). 
These are more than merely model agnostic, since they have the potential to be sensitive to a much broader range of new physics scenarios, in particular because the DNN finds it own solutions to the problems posed to it. 
It may find patterns in the high-dimensional data that physicists have not thought to look for and identify relevant high-level features of the data that physicists have not focused on. 
Model-agnostic approaches, such as simplified models, by contrast may only be sensitive to a small range of BSM scenarios. 
Rather than a single target model, simplified models have many, which has some benefits, but it is not the same as having no target model.

These two necessary conditions may seem to be in contrast to the description of model independence as a spectrum. 
Recall first that model-independent searches are at one end of a range of searches varying in model dependence.
Further, it is itself not a hard and fast category. 
Many different kinds of searches and strategies have a well-defined background and no target model and some of these still feature roles for modelling assumptions, such as in BSM benchmark testing. 
Reducing the role of modelling assumptions is then more of a desideratum over and above the threshold imposed by the two necessary conditions. 
Fulfilling this desideratum is not easily accomplished, but minimizing the compromises that come with this is a major task for future DL projects in HEP. 

Some, such as \citep{mccoymassimi}, have applied the term \textit{model independent} to searches that still have target models (albeit not a single one). 
Others, such as \citep{Dillon:2023zac} use the term \textit{model agnostic} to searches that do not have target models, but that make use of certain modelling assumptions. 
The schema I propose makes a case for a clear and precise use of the term: it provides clear conditions under which something is model independent, and it makes sense of the variety of approaches by allowing that it is a part of spectrum. 
The spectrum of model dependence ranges from having one or a few target models, to having many target models, to having none. 
And since the characteristic aim of model independence is to reduce the reliance on modelling assumptions in order to broaden the realm of possible discoveries, there are also different roles of different strengths that modelling assumptions can reasonably play in searches that are properly model independent. 

Of course, one should be wary of claims that DL anomaly searches can capture any deviations whatsoever. 
They have really only been tested against particular BSM scenarios. 
One may argue that the network has only learned the background well enough to distinguish some BSM scenarios and where it has not been tested, one should not assume it will work.
While this should temper our excitement, it has been demonstrated in various anomaly detection studies and competitions that a network tested to have high anomaly scores on various BSM scenarios, is able to flag further kinds of new signals as anonymous. 
In essence, this is little other than generalizability. 
DL models in general have only been tested on particular datasets, but they can perform extraordinarily well on new data they have never encountered. 
This gives credence to the idea that the network has really learned something integral to training data---in our case that the AEN has actually learned relevant parts of the SM. 

Interpretation after the fact is also not trivial. 
The best that can be done in the programme for anomaly detection is to model the background as well as possible, test that a network flags known BSM scenarios, expand these scenarios as wide as possible, and continue to test if it can flag anomalous inputs in real data for further investigation. 
Of course, one cannot validate the network on discovering unexpected discoveries, but if we have good reasons to believe that DL models can generalize, then these anomaly detection networks can be very broadly applicable for discovering new physics.

\section{Concluding Remarks}

I have argued in this paper that the term \textit{model independence} is an important and meaningful one and I have proposed a schema to make sense of the broad range of uses. 
Model independence marks a third, often under-represented advantage of DL: not only can DL demonstrate improved performance over traditional methods and increase the computational efficiency, it can also provide a less biased way of searching for new physics and thus opens an avenue for potentially dissolving the current model underdetermination in particle physics. 
Sensitivity towards a broader range of anomalies carries certain practical difficulties, but a careful look at physics practice and a view of model independence that is situated along a spectrum shows that DL is major avenue for model independence. 
DNNs are powerful tools for discovering patterns in high-dimensional data, but much work remains to be done to get as much as use out of them as possible.

DNNs will not replace physics models or top-down model testing. 
The unsupervised discovery of an anomaly with a VAE will not provide physicists with a model of new physics. 
Rather, it flags the data for further study to corroborate the anomaly, to model the phenomena, and to test those physics models. 
Model-independence is a part of a multi-pronged approach to finding new physics, but one with an important role in filling the gaps of traditional model-based searches.

\bibliography{../../library}
\bibliographystyle{apalike}

\end{document}